\documentclass[9pt,twocolumn,twoside]{opticajnl}
\journal{opticajournal} % use for journal or Optica Open submissions

\setboolean{shortarticle}{false}
\setboolean{displaycopyright}{false} % Do not include copyright or licensing information in submission.

\title{Anomalous Purcell decay of strongly driven inhomogeneous emitters coupled to a cavity}

\author[1,2,3]{Michael T. Solomon}
\author[1]{Martin Koppenh\"ofer}
\author[1]{Mikhail Mamaev}
\author[1,2]{Cheng Ji} 
\author[1,2]{Gregory Grant} 
\author[1,2]{Ignas Masiulionis} 
\author[2,3,7]{Sean E. Sullivan} 
\author[2,3,1]{F. Joseph Heremans}
\author[1,2,3]{Supratik Guha}
\author[1,2,3,4]{David D. Awschalom}
\author[1,2,3]{Aashish A. Clerk}
\author[5,6,3,*]{Alan M. Dibos}

\affil[1]{Pritzker School of Molecular Engineering, University of Chicago, Chicago, IL 60637, USA}
\affil[2]{Materials Science Division, Argonne National Laboratory, Lemont, IL 60439, USA}
\affil[3]{Center for Molecular Engineering, Argonne National Laboratory, Lemont, IL 60439, USA}
\affil[4]{Department of Physics, University of Chicago, Chicago, IL 60637, USA}
\affil[5]{Nanoscience and Technology Division, Argonne National Laboratory, Lemont, IL 60439, USA}
\affil[6]{Center for Nanoscale Materials, Argonne National Laboratory, Lemont, IL 60439, USA}
\affil[7]{Current Address: memQ Inc., Chicago, IL 60615, USA}
\affil[*]{adibos@anl.gov} %% email address is required; see note below about the corresponding author designation

\begin{abstract} 
We perform resonant fluorescence lifetime measurements on a nanocavity-coupled erbium ensemble as a function of cavity-laser detuning and pump power. 
Our measurements reveal an anomalous three-fold suppression of the ensemble Purcell factor at zero cavity detuning and high pump fluence. 
We capture qualitative aspects of this decay rate suppression using a 
Tavis-Cummings model of non-interacting spins coupled to a common cavity.
\end{abstract}

\usepackage[version=3]{mhchem}
\usepackage[normalem]{ulem}

\begin{document}

\maketitle

\newcommand{\Martin}[1]{\textcolor{orange}{(MK: #1)}}
\newcommand{\Alan}[1]{\textcolor{red}{(AD: #1)}}
\newcommand{\Mike}[1]{\textcolor{magenta}{(MS: #1)}}
\newcommand{\Aash}[1]{\textcolor{brown}{(AC: #1)}}

%%% the two-column template has an ``intelligen'' eqref command which tries to add Eq(s). before the
%%% reference. This is incompatible with our writing style, so let's deactivate it.
\renewcommand{\eqref}[1]{(\ref{#1})}

%%%%%%%%%%%%%%%%%%%%%%%%%%  body  %%%%%%%%%%%%%%%%%%%%%%%%%%
\section{Introduction}

Tailoring the interaction between atoms and their electromagnetic environment is of both fundamental interest and practical relevance, e.g., for applications in quantum communication and quantum information processing \cite{Hammerer2010,Reiserer2015}. By tuning the photon density of states, one can drastically modify the emission properties of atoms \cite{Purcell1946}, a phenomenon that underpins the thriving research area of cavity quantum electrodynamics (cQED) \cite{Walther2006}. Prototypical cases  are the interaction between a single two-level system (which we will interchangeably refer to as a ``spin'') and a single mode of a radiation field \cite{Jaynes1963}, the collective interaction of atoms with their electromagnetic environment leading to the effects of super- and subradiance \cite{Dicke1954,GrossHaroche1982,Temnov2005}, as well as photon-mediated collective interactions \cite{Norcia2018}, and large ensembles of weakly excited spins with identical or inhomogeneously broadened transition frequencies \cite{Jaynes1963,Kurucz2011}. The limit where the spins are only weakly perturbed away from their ground or excited states is appealing, since the spin dynamics become linear and analytical studies of the dynamics become possible \cite{Julsgaard2012,Julsgaard2013,Blaha2022}. Much less is known, however, in the case of highly excited spins where the intrinsic nonlinearity of the spin dynamics due to their finite-dimensional Hilbert space needs to be taken into account.

Trapped atomic systems are the prototypical spins for investigating optical relaxation dynamics because of their coherence and homogeneity in ensembles. In cQED, there is particular interest in modifying the atom-photon coupling strength through the use of high-quality-factor (Q) and small-mode-volume optical cavities. While significant progress has been made in integrating conventional trapped atomic systems with these types of resonators \cite{Thompson2013,  Gallego2018, Kumar2023}, it still remains an outstanding technical challenge to experimentally achieve high coupling with modest ensembles of atoms to probe the rich phenomena available from collective effects \cite{Mabuchi1999, Goban2015, Chang2018, Mirhosseini2019}. In contrast, rare-earth ions embedded in solid-state hosts and integrated with nanoscale optical cavities\textemdash{which are being explored for use in quantum memories \cite{Kindem2020, Raha2020, Chen2020, Ourari2023} and microwave-to-optical transduction \cite{Xie2021, Rochman2023}, have naturally long optical lifetimes that are ideally suited for optical-decay-modification experiments. These features of nanocavity-coupled rare-earths are typically at the expense of large inhomogeneous broadening, but this trait has recently been shown to be an important factor in the observation of collectively induced transparency \cite{Lei2023}, where the dynamics of a strongly driven and disordered ensemble can modify the steady-state system response due to an interplay of quantum interference and collective effects. Similarly in this work, we study an ensemble of rare-earth ions coupled to a relatively high-Q, small-mode-volume optical cavity and subjected to a resonant laser driving field. The large inhomogeneous linewidth of the ensemble, and the relatively narrow resonance of the cavity, enable us to explore surprising effects that manifest themselves in the dependence of the highly excited spin ensemble's optical relaxation rate on both cavity-laser detuning and optical drive power.  These occur in a regime where the collective cooperativity is likely large and disorder plays a non-trivial role.  
%how the optical relaxation of the highly excited spins\textemdash varies as a function of both cavity-laser detuning and optical drive power\textemdash likely a regime where the collective cooperativity is large and disorder plays a non-trivial role.

Our specific spin-cavity system consists of trivalent erbium ions (Er$^{3+}$ or Er for brevity) incorporated into a thin layer of TiO$_2$ grown atop silicon-on-insulator wafers \cite{Dibos2022, Ji2023}. The Er ions are evanescently coupled to a 1D photonic crystal cavity that is patterned through both the TiO$_2$ and Si device layers. In a prior study, we attempted to measure the Purcell enhancement in Er-doped TiO$_2$ films coupled to a nanophotonic cavity as we tuned the cavity resonance across a particular optical transition \cite{Dibos2022}. During the course of those experiments, we noticed features resembling the saturation of the Purcell enhancement at high laser pump powers. However, the overall photon collection efficiency and fluorescence signal were too low to perform sufficient sweeps of the optical drive. Here, the Er concentration is three-fold increased by growing more highly doped films and the device-to-fiber coupling efficiency is five-fold increased by incorporating an inverse taper in-coupling structure. We report anomalously slow relaxation rates in the temporal response of highly-excited and inhomogeneously broadened spin ensembles. Finally, we develop a simple theoretical model to explain the qualitative features in the Purcell-enhanced ensemble decay. We reproduce qualitative features of this decay in semiclassical simulations of a Tavis-Cummings model of non-interacting spins coupled to a common cavity. Finally, we derive a simple theoretical model that provides physical intuition for this phenomenon.

\section{Experimental details and results}

\subsection{Samples and device fabrication}

 For this study, we use molecular beam deposited Er-doped TiO$_2$ films that contain both rutile and anatase crystalline grains, as was seen in prior work \cite{Dibos2022}. Previous experiments have shown that the inclusion of undoped spacer layers surrounding a doped region can reduce optical inhomogeneous broadening for epitaxial Er:Y$_2$O$_3$ \cite{singh2020} and Er:TiO$_2$ thin films \cite{Shin2022}, as well as spectral diffusion and inhomogeneous linewidth broadening for Er in polycrystalline TiO$_2$ films \cite{singh2022}. However, a thick TiO$_2$ layer is more difficult to etch anisotropically into high Q cavities. With both of these considerations in mind, we employ a `$10/10/10$' heterostructure: a 10 nm undoped TiO$_2$ buffer layer at the Si interface, a 10 nm Er-doped TiO$_2$ layer in the middle, and a 10 nm undoped TiO$_2$ capping layer on top. The sample is grown on a standard SOI wafer used for silicon photonics. During the molecular beam deposition of Er:TiO$_2$, we use a substrate temperature of 390 $^{\circ}$C and metallic Er cell temperature of 900 $^{\circ}$C. This Er temperature corresponds to flux that will yield 120 (130) ppm Er if the crystal phase of the TiO$_2$ is rutile (anatase) due to the difference in density.

\begin{figure*}
    \centering
    \includegraphics[width=0.8\textwidth]{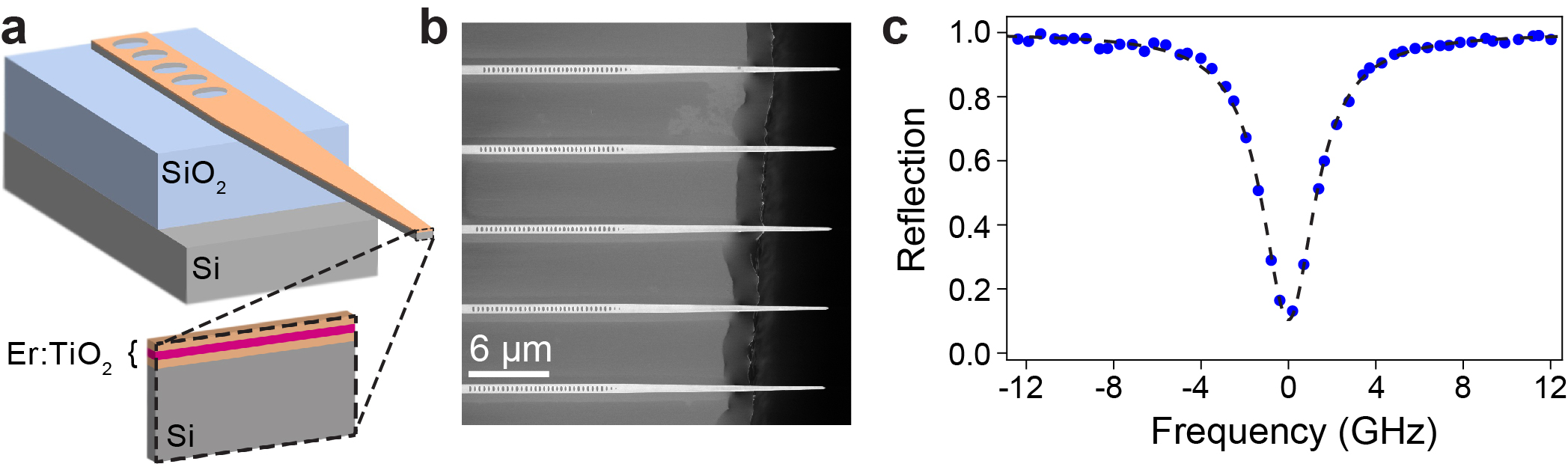}
    % \captionsetup{width=\textwidth}
    \caption{Er-doped TiO$_2$-Si 1D photonic crystal cavities. (a) Schematic of fabricated devices showing the inverse taper of the waveguide extending off the chip. The inset at the end facet highlights the doped TiO$_2$ heterostructure deposited on top of the Si device layer consisting of 10 nm of undoped TiO$_2$/10 nm doped TiO$_2$ ($\sim120$ ppm Er, colored red)/10 nm undoped TiO$_2$. (b) Scanning electron microscope image of the nanophotonic cavity composed of elliptically shaped holes etched through the TiO$_2$/Si device layers and the waveguide inverse taper protruding from the chip edge. (c) Normalized laser reflection spectrum of the cavity tuned onto resonance with the $Z_{1}\rightarrow Y_{1}$ optical transition for Er in rutile phase TiO$_2$ ($\lambda = 1520.52$ nm) at a temperature of 3.5 K. Using a Lorentzian fit (dashed line), the cavity quality factor gives Q=$(6.71 \pm 0.11) \times 10^4$.}
    \label{fig:1}
\end{figure*}

Our Er:TiO$_2$/Si photonic crystal cavities are optically addressed in a one-sided configuration via lensed optical fiber. The waveguide width and photonic crystal cavity design parameters are identical to that shown previously \cite{Dibos2022}. However, for this work we wanted to improve the coupling efficiency between the lensed fiber and the cavity, as previous devices involved end-fire coupling into a cleaved waveguide with an efficiency of $10-15\%$. To do so, we used Lumerical 3D FDTD simulations to optimize the design of an inverse taper that extends off the edge of the chip (Fig.~\ref{fig:1}a) for better mode matching to the lensed fiber \cite{Meenehan2014,Dibos2018}, and simulation details are provided in the Supplementary Information (SI). The simulations show that an adiabatic reduction in waveguide width from 650 nm (where the cavity holes end) down to 200 nm over a 14 $\mu$m length, allows for a fiber coupling efficiency up to $80\%$. 

We fabricate these Er:TiO$_2$ on Si 1D photonic crystal cavities as outlined previously\cite{Dibos2022}, but with additional steps to selectively undercut the inverse tapers without damaging the Si and TiO$_2$ (complete fabrication flow is provided in the SI, Fig.~S2a). We deposit a device protection layer of conformal thermal ALD alumina, which covers the top surface of the sample as well as the exposed sidewalls and end facet of the inverse taper. We then cleave the SOI chip adjacent to the termini of the tapers via a precision cleaver under an optical microscope. After this, we perform an isotropic etch of the Si handle layer via XeF$_2$, followed by vapor HF (VHF) etching to undercut the buried SiO$_2$. After sufficient VHF exposure to fully remove the 2 $\mu$m buried oxide layer underneath the suspended portion of the taper, we strip the protective alumina layer using a wet etchant and anneal the sample. SEM and optical images of the final devices are shown in Figures \ref{fig:1}b and S2b (SI), respectively. It should be noted that as long as the vapor processes yield a lateral undercut length of at least 4 $\mu$m of the 14 $\mu$m taper, the mode is sufficiently waveguided such that minimal light is lost into the substrate (Fig.~S1a, SI).

\subsection{Optical measurements at 3.5 K}

We perform our optical device measurements in a fiber-accessible cryostat at a sample temperature of 3.5 K, and the experimental configuration is discussed elsewhere \cite{Dibos2022}. For this work, we will explore ensemble cavity coupling to the $Z_{1}\rightarrow Y_{1}$ optical transition for Er in rutile phase TiO$_2$ ($\lambda = 1520.52$ nm) \cite{Phenicie2019}. We then use a continuous-wave (CW) tunable laser (Toptica CTL 1550) to sweep through the device resonance and measure the reflection spectrum for the cavity. The cavity full-width half-maximum (FWHM) linewidth is $2.94 \pm 0.05$ GHz, when centered at $\lambda = 1520.52$ nm, and this corresponds to a Q =$(6.71 \pm 0.11) \times 10^4$ (Fig.~\ref{fig:1}c) with a waveguide-cavity mode coupling efficiency, $\eta$ = 34\%, yielding a decay rate from the cavity of $\kappa_\mathrm{c} = 0.34 \kappa$. In a simplified ambient setup with a closed-loop nanopositioner, we measure our suspended in-coupling structures to have a typical lensed fiber-to-waveguide coupling efficiency of $\approx75\%$, which is close to our simulated maximum of $80\%$. However, when we perform the same characterization in the cryostat at 3.5 K, we see values closer to $\approx65\%$.

\begin{figure*}
    \centering
    \includegraphics[width=0.8\textwidth]{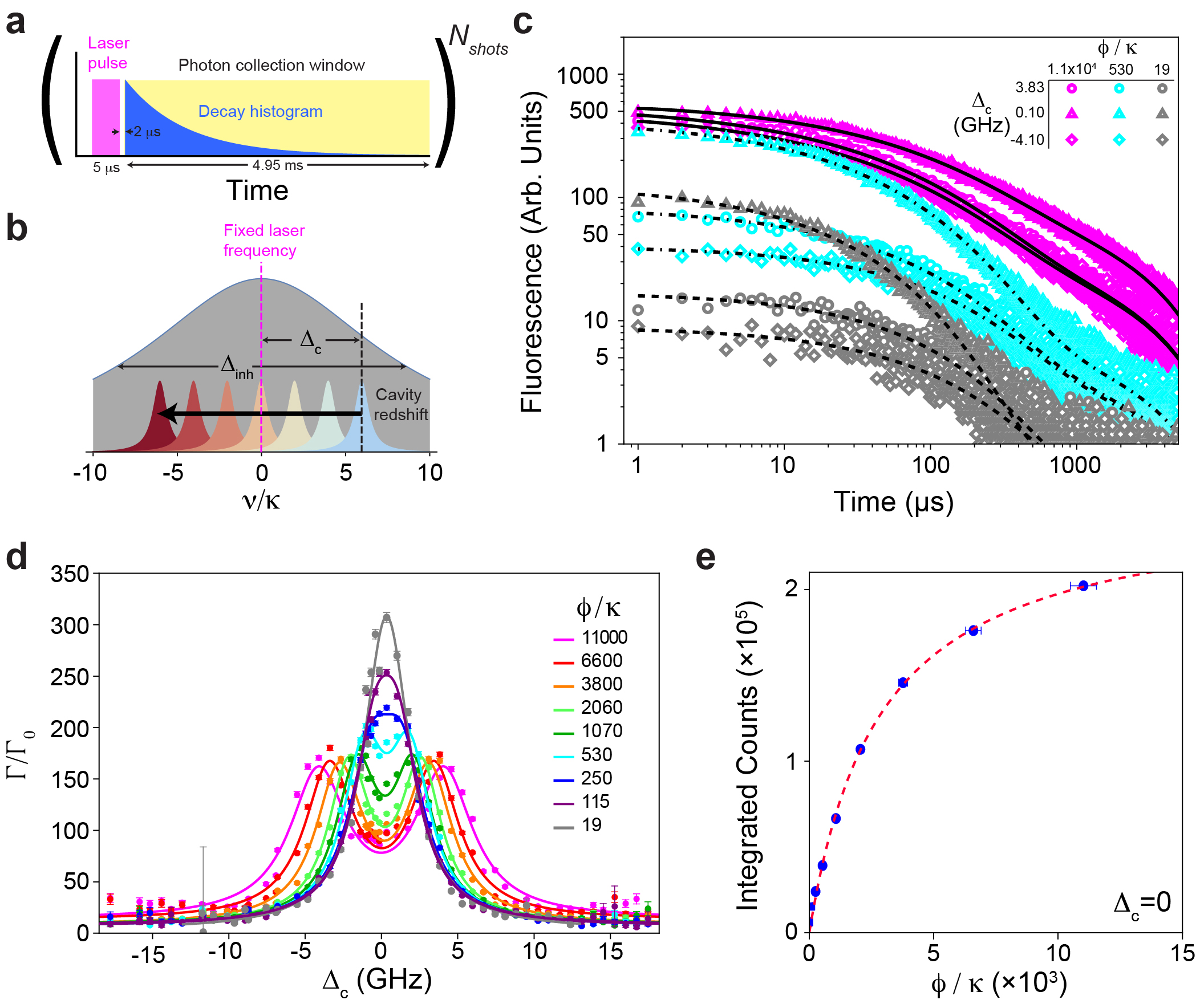}
    \caption{
    % \Martin{@Alan: Do you want to include the PLE fit into the plot for clarity? I'll also add the corresponding plot to Fig.~\ref{fig:3} with the theory PLE fit. Can also put them in the SM if it's too much work now...}
    PLE Measurements on cavity-coupled Er ensembles at 3.5 K. (a) Schematic of our photoluminescence excitation measurement sequence. We first generate a 5 $\mu$s long laser excitation pulse (magenta). After a 2 $\mu$s wait time, we open an additional modulator in the collection path to start the photon collection window (yellow) to enable detection of the fluorescence from the ensemble. This single pulse sequence is repeated over $N_\mathrm{shots}$ measurements to build up a histogram of the decay (blue). (b) Schematic of the cavity tuning mechanism in the measurement. For all of these measurements the laser is locked to a particular frequency at the center of the ensemble's inhomogeneous distribution. This distribution is broad relative to the cavity linewidth ($\kappa$) and is represented by the wide gray Lorentzian ($\approx 17 \kappa$). The cavity is slowly redshifted through the resonance through gas condensation tuning, and its frequency relative to the laser is given by $\Delta_\mathrm{c}$. The experimental extent of $\Delta_\mathrm{c}$ is $\approx 12 \kappa$ as represented with the blue-to-red transition of the cavity Lorentzian. (c) Nine example experimental time traces for three different incident photon flux values ($\phi$) and cavity-laser detuning values ($\Delta_\mathrm{c}$). The various black lines are fits to the experimental data as described in the main text. (d) The spontaneous decay rate enhancement ($\Gamma / \Gamma_0$) as a function of the cavity-laser detuning ($\Delta_\mathrm{c}$) for increasing photon fluxes ($\phi$). The $\Gamma / \Gamma_0$ values are extracted from the stretched exponential time constants fit to experimental data such as those in (c). The solid lines for each flux are from a pair of Lorentzian lineshapes fit to the experimental data. (e) Measurements of the integrated fluorescence intensity when the cavity is resonant with the optical transition ($\Delta_\mathrm{c}$ = 0.10 GHz), the red dashed line is a saturation fit as described in the main text.}
    \label{fig:2}
\end{figure*}

We explore the cavity-coupled optical relaxation dynamics in these ensembles using resonant photoluminescence excitation (PLE) measurements as depicted in Figure \ref{fig:2}a. We use three acousto-optic modulators (AOM) in series to generate 5 $\mu$s laser pulses from our tunable CW laser. After the end of the laser pulse, we wait 2 $\mu$s before opening a fourth AOM in the collection path to our superconducting nanowire single-photon detectors (SNSPDs), which are not gated. We can repeat each pulse sequence many times (typically $N_\mathrm{shots}=4000$) to create a histogram of the ensemble fluorescence lifetime and measure the integrated intensity. Using this approach, resonant PLE from a laser sweep near 1520.52 nm, yields a broad Er$^{3+}$ ensemble linewidth of 50 GHz, as was shown previously \cite{Dibos2022}, and is represented by the wide gray Lorentzian shown in Figure \ref{fig:2}b. The inhomogeneous linewidth $\Delta_\mathrm{inh}$ is much broader than the cavity linewidth ($\kappa$=2.94 $\pm$ 0.05 GHz), and we use nitrogen gas condensation on the cavity to tune the refractive index and slowly redshift the cavity resonance through the center of the inhomogeneous linewidth. For the majority of measurements, the laser is locked at $\lambda = 1520.52$ nm using a wavemeter (WS-8-10, High Finesse) calibrated against a fixed frequency reference laser (SLR-1532, High Finesse). We periodically interrupt PLE measurements to perform CW laser reflection measurements, such as the one shown in Figure~\ref{fig:1}c, to precisely measure the cavity-laser detuning parameter, $\Delta_\mathrm{c}$. Given that the inhomogeneous linewidth of the $Z_{1}\rightarrow Y_{1}$ transition for Er in rutile TiO$_2$ is broad, it is important to investigate if the homogeneous linewidth of these emitters is sufficiently narrow to enable cavity-based lifetime enhancement (`bad cavity' limit). Similar to previous work \cite{Dibos2022}, we have performed transient spectral hole burning (TSHB) measurements at 3.5 K, to find an upper bound on the homogeneous and spectral diffusion linewidths \cite{Weiss2021}, as shown in Figure S3 of the SI. We can see that for the range of relevant incident photon fluxes for which we see TSHB contrast, the TSHB linewidth is $\leq 0.3$ GHz, and we assume that for even higher fluxes our cavity-coupled ensemble remains in the `bad cavity' regime.

We can now perform lifetime measurements at a variety of input laser powers for each cavity detuning. For later comparison with theory, we quantify the incident laser brightness in terms of $\phi / \kappa$, the photon flux given by the number photons of incident on the cavity per cavity lifetime (1/$\kappa$). 
%\Aash{This definition seems weird, shouldn't it be per second?}
%\Martin{@Aash: For the theory section, the cavity lifetime is a free parameter. We could fix it to be the same as the experimental value and then specify everything in photons/second, but for the sake of easy comparison with a broad range of experiments I would prefer to normalize by the cavity decay time.}
%\Martin{The original wording was a bit unclear. $\phi$ as defined below in the theory section is a flux, i.e., it should have units $1/time$ and we should indicate this on the plot axes and in the legends. The way things are plotted right now, we suggest $\phi$ is dimensionless (i.e., a flux times $1/\kappa$) but then ``photon flux incident on the cavity per cavity lifetime'' should be replaced by ``photon flux incident on the cavity integrate over a cavity lifetime'' (@Alan which is maybe indeed what we calculated for the axes?).} 
%\Martin{@Alan: we need to add $\kappa$ to the plot legends and axes or plot $\phi/\kappa$.}
It is important to note that for a given detuning we perform a full sweep of the incident fluxes and then measure the cavity reflection to gauge the resonance position. The cavity tuning rate is sufficiently slow to enable us to average our resultant lifetime histograms over five power sweeps, as this is useful in achieving sufficient detected photon numbers for the lower flux measurements, as well as averaging over fluctuations in the total intensity, as will be discussed later. As shown in Figure \ref{fig:2}c, for the lowest laser power used ($\phi= 19 \kappa$) we measure a fast decay time close to resonance ($\Delta_\mathrm{c} \approx 0$) and slower decay at modest cavity detunings of $\approx \pm4$ GHz. In contrast, even though the fluorescence intensity is brighter for the highest laser power ($\phi = 1.1 \times 10^4 \kappa$), we see faster decay when the cavity is detuned from the transition than on resonance. Following previous work \cite{Dibos2022,Lei2023}, we can fit the experimental data using a function of the form: $A\exp[-(t/\tau_1)^d] + B\exp(-t/\tau_2) + C$. For our case, the stretched exponential with lifetime $\tau_1$ represents the fastest ensemble fluorescence decay rate ($\Gamma = 1/\tau_1$) mediated by coupling to the cavity. The stretching exponent ($d$) attempts to capture the distribution of individual ion-cavity coupling strengths ($g$) within the ensemble, and the single exponential decay constant represents the much slower decay from Er ions along the inverse taper waveguide because the ions are located everywhere in the film. It is important to note that for these measurements, the specific fit for $\tau_1$ is robust regardless of the other parameters ($A$, $B$, $d$, $\tau_2$, and $C$) \cite{Dibos2022}. If we also measure the optical lifetime of the fluorescence decay rate in a bare waveguide without a cavity $\Gamma_0 = 1 / \text{T}_\text{1}$, where T$_\text{1} = 5.370 \pm 0.013$ ms (Fig.~S4, SI), then the cavity-mediated Purcell enhancement, $\Gamma /\Gamma_0$, can be computed for each incident flux and cavity detuning, and the resultant plot is shown in Figure \ref{fig:2}d.
% and at the lowest powers we see a Lorentzian lineshape in the Purcell factor with a FWHM linewidth of $3.96 \pm 0.24$ that is slightly larger but similar to the cavity linewidth, as was seen previously\cite{Dibos2022}.
Most interesting is that at higher incident fluxes we see a double-peak decay profile appear, and the decay rate appears to slow when the cavity is resonant with the driving field. This double-peaked detuning profile is significantly wider than the lower power cases: a single Lorentzian fit of the $\phi=19 \kappa$ case yields a FWHM = $3.96 \pm 0.24$ GHz, which is 35\% wider than the cavity linewidth measured via reflection. However, if we integrate all photons detected at a particular detuning for each flux (regardless of time bin), the commensurate normalized intensity versus detuning plot does not show this double-lobed structure (Fig.~S5, SI). Instead, all curves are qualitatively similar with comparable widths (FHWM $\sim 5$ GHz). It is important to note that we observe this power dependence in all devices that can be tuned through the rutile Er transition, and if we probe cavities that are tuned instead to the Er$^{3+}$ in anatase TiO$_2$ transition (1532.6 nm), we also see very similar qualitative behavior of the lifetime decay versus detuning (Fig. S6, SI) as that shown in Figure ~\ref{fig:2}d.

When we plot the total PLE intensity for the resonant case, we can see the onset of saturation (Fig.~\ref{fig:2}e), where the dashed red line is a fit to a standard photoluminescence-excitation (PLE) function of the form 
\begin{align}
    P(\phi) = \frac{p_1}{p_2 + 1/(\phi / \kappa)}~.
    \label{eq:PLE}
\end{align}
%If we also integrate over all detunings, we can see a similar onset of saturation (Fig.~\ref{fig:2}e, right axis). 
If we use the total experimental photon collection efficiency of our system (0.023) and the detected number of photons, we can infer $\sim 2200$ photons generated in the cavity per pulse at the highest pump power ($\phi=1.1 \times 10^4 \kappa$). 
Since we are in a saturation regime, we can generally conclude that the number of ions that are addressed with our optical pulse is close to this value. 
Using our device geometry and doping density, we very coarsely estimate that there are $7 \times 10^4$ total ions along the cavity, though our uncertainty is quite large given the unknown proportion of Er ions residing in the rutile vs anatase grains. However, if we use this value and make another coarse estimate that if the spectral window that is being addressed with these highest power pulses is approximately the cavity linewidth ($\kappa=2.94$ GHz), then we expect $\sim 2600$ ions to couple. Since our estimates on ion number and photons generated in the cavity are reasonably close (in a regime where the ensemble appears to be saturating) we expect that the number of participating ions is close to 2000, and this estimate is useful in conducting simulations to explain the features in the Purcell factor detuning dependence.

\section{Theory and discussion}

\subsection{Semiclassical model}

To model the interaction between the nanophotonic cavity and the Er ions coupled to it, we treat each ion as a two-level system, representing the $Z_1$ and $Y_1$ levels, which is interacting with the cavity mode via a standard Tavis-Cummings interaction, 
\begin{align}
    \hat{H} = \Delta_\mathrm{c} \hat{a}^\dagger \hat{a} + \sum_{j=1}^N \delta_j \frac{\hat{\sigma}_z}{2} + \sum_{j=1}^N g_j \left(\hat{a}^\dagger \hat{\sigma}_-^j + \hat{a} \hat{\sigma}_+^j \right) - i \sqrt{\kappa_\mathrm{c}} \beta_\mathrm{in} \left( \hat{a}^\dagger - \hat{a} \right)~.
    \label{eq:H_TC}
\end{align}
Here, $\hat{\sigma}_z^j$ is the Pauli $z$ matrix of ion $j$, $\hat{\sigma}_\pm^j$ are the associated spin raising and lowering operators, respectively, and $\hat{a}$ is the lowering operator of the cavity mode. 
The cavity is driven by a laser at frequency $\omega_\mathrm{dr}$ which generates an input photon flux $\phi = \vert \beta_\mathrm{in} \vert^2$.
We work in a frame rotating at the drive frequency and $\Delta_\mathrm{c} = \omega_\mathrm{c} - \omega_\mathrm{dr}$ ($\delta_j = \omega_j - \omega_\mathrm{dr}$) is the corresponding detuning of the cavity (ion $j$). 
Relaxation of each ion and damping of the cavity are described by the quantum master equation
\begin{align}
    \frac{\mathrm{d}}{\mathrm{d} t} \hat{\rho} = - i \left[ \hat{H}, \hat{\rho} \right] + \kappa \mathcal{D}[\hat{a}] \hat{\rho} + \gamma \sum_{j=1}^N \mathcal{D}[\hat{\sigma}_-^j] \hat{\rho} ~,
    \label{eq:QME}
\end{align}
where $\mathcal{D}[\hat{o}] \hat{\rho} = \hat{o} \hat{\rho} \hat{o}^\dagger - \left\lbrace \hat{o}^\dagger \hat{o}, \hat{\rho} \right\rbrace/2$ is a Lindblad dissipator. 
Here, $\gamma$ is the relaxation rate of the ions and $\kappa$ is the total cavity damping rate, which accounts for the damping rate $\kappa_\mathrm{c}$ due to the input port (through which the drive laser enters the cavity) and other internal losses.

Since it is intractable to solve the quantum master equation~\eqref{eq:QME} for large ensembles of ions, we derive equations of motion for the coherent cavity field $a = \langle \hat{a} \rangle$ as well as the spin populations $s_z^j = \langle \hat{\sigma}_z^j \rangle$ and coherences $s_-^j = \langle \hat{\sigma}_-^j \rangle$. 
Because of the nonlinearity of the spins, these equations of motion will not form a closed set of differential equations.
We therefore perform a semiclassical approximation $\langle \hat{o}_1 \hat{o}_2 \rangle \approx \langle \hat{o}_1 \rangle \langle \hat{o}_2 \rangle$ \cite{WallsMilburn1994}, also known as a first-order mean-field approximation, which leads to
\begin{align}
    \frac{\mathrm{d}}{\mathrm{d} t} a 
        &= - \left( i \Delta_\mathrm{c} + \frac{\kappa}{2} \right) a - i \sum_{j=1}^N g_j s_-^j - \sqrt{\kappa_\mathrm{c}} \beta_\mathrm{in} ~, 
        \label{eq:SC:a}\\
    \frac{\mathrm{d}}{\mathrm{d} t} s_-^j 
        &= - \left( i \delta_j + \frac{\gamma}{2} \right) s_-^j + i g_j a s_z^j ~, 
        \label{eq:SC:sm}\\
    \frac{\mathrm{d}}{\mathrm{d} t} s_z^j
        &= 2 i g_j \left[ a^* s_-^j - a (s_-^j)^* \right] - \gamma (1 + s_z^j)~.
        \label{eq:SC:sz}
\end{align}
Even though this approximation neglects all correlations between the cavity field and the spins and does not take into account the presence of the ions coupled to the inverse taper, the resulting equations of motion are still able to capture the relevant qualitative aspects of the experimental data, as we will show below.

To simulate the dynamics of a disordered ensemble of ions, we randomly draw the detunings $\delta_j$ of $N-1$ ions from a Lorentzian distribution $L(\delta, \tilde{\delta}, h) = h / \pi [(\delta - \tilde{\delta})^2 + h^2]$, centered around $\tilde{\delta} = 0$ with a full-width-at-half-max (FWHM) of $2 h = \Delta_\mathrm{inh}$.
The remaining last ion is always chosen to be resonant with the drive. 
Similarly, we randomly draw the coupling strengths $g_j$ of each ion from a Gauss distribution $G(g,\overline{g},\sigma) = \exp[- (g-\overline{g})^2/2 \sigma^2] / \sqrt{2 \pi} \sigma$ with mean coupling strength $\overline{g}$ and standard deviation $\sigma$. 
This procedure ensures that the detunings $\delta_j$ and the couplings $g_j$ of each ion are uncorrelated random variables. 
Since a finite number of spins is in general not sufficient to smoothly sample a broad inhomogeneous frequency distribution, we repeat this procedure $n_\mathrm{traj}$ times to generate a set of random disorder realizations.
For each of these disorder realizations, we initialize the cavity in the va\-cu\-um state and the ions in the ground state, i.e., $a=s_-^j=0$ and $s_z^j=-1$, and evolve the semiclassical equations of motion~\eqref{eq:SC:a} to~\eqref{eq:SC:sz} with the laser drive switched on, $\beta_\mathrm{in} > 0$, for a time $t_\mathrm{pulse}$.
Subsequently, we use the final state at $t_\mathrm{pulse}$ as the new initial state for a simulation of the fluorescence of the ion ensemble, where the laser drive has been switched off, $\beta_\mathrm{in} = 0$.
The semiclassical output field of the cavity is $a_\mathrm{out} = \sqrt{\kappa_\mathrm{c}} a + \beta_\mathrm{in}$, and the photon flux impinging on a photodetector measuring the output mode is $\vert a_\mathrm{out} \vert^2$.

For the numerical simulations, we choose parameters that are inspired by the corresponding experimental values but allow us to perform efficient simulations of the inhomogeneously broadened spin ensemble. 
Based on the transient-spectral-hole-burning experiments, we choose the spin decay rate to be $\gamma = 0.005\kappa$. 
In order to get a larger output mode amplitude, we increase the output coupling strength compared to the experimental value and use $\kappa_\mathrm{c} = 0.8 \kappa$.
We reduce the width of the inhomogeneous distribution to $\Delta_\mathrm{inh} = 5 \kappa$ to be able to perform disorder averages using $N=61$ ions and $n_\mathrm{traj} = 120$ random disorder realizations. 
Since the experimental value of the coupling strength of the ions is unknown, we choose $\overline{g} = 0.07 \kappa$ and $\sigma = 0.01 \kappa$, such that $C = 4 N \overline{g}^2 / \kappa \gamma \gg 1$ should allow us to resolve collective effects in the simulations.

\subsection{Explicit form of the state prepared by excitation pulse}

The state of the ion ensemble at the end of the excitation pulse can be well understood using an argument that combines single-spin (i.e., local) dynamics and the concept of collectively induced transparency (CIT), which has recently been introduced and experimentally demonstrated \cite{Lei2023}.
For an excitation pulse that is long compared to both the cavity decay timescale $1/\kappa$ and the intrinsic spin decay timescale $1/\gamma$, the system will approach a drive-dependent steady state. 
After eliminating the spin coherence $s_-^j$ by solving for the steady state of Eq.~\eqref{eq:SC:sm}, one finds that the steady-state cavity field and the spin excitations are given by the self-consistent solution of 
\begin{align}
    a_\mathrm{ss} &= \frac{i \sqrt{\kappa_\mathrm{c}} \beta_\mathrm{in}}{\Delta_\mathrm{c} - i \frac{\kappa}{2} + \sum_{j=1}^N \frac{g_j^2}{\delta_j - i \gamma/2} s_{z,\mathrm{ss}}^j} ~, 
    \label{eq:ass}\\
    s_{z,\mathrm{ss}}^j &= - \left( 1 + \frac{2 g_j^2}{\delta_j^2 + \gamma^2/4} \vert a_\mathrm{ss} \vert^2 \right)^{-1} ~.
    \label{eq:szss}
\end{align}
The last term in the denominator of Eq.~\eqref{eq:ass} represents a spin-excitation-dependent self energy that can renormalize both the effective detuning $\Delta_\mathrm{c}$ and the decay rate $\kappa$ of the cavity. 
%cavity mode acquires a spin-excitation-dependent self-energy term,
%\begin{align}
%    \frac{\mathrm{d}}{\mathrm{d} t} a
%        &= - \left( i \Delta_\mathrm{c} + \frac{\kappa}{2} - \sum_{j=1}^N \frac{g_j^2 s_z^j}{i \delta_j + \gamma/2} \right) a - \sqrt{\kappa_\mathrm{c}} \beta_\mathrm{in} ~,
%\end{align}
%which is given by the last term in the parentheses. 
In the limit of weakly excited spins, $s_z^j \approx -1$, Eq.~\eqref{eq:ass} can be averaged over the disorder distributions $L(\delta,0,\Delta_\mathrm{inh}/2)$ and $G(g, \overline{g}, \sigma)$,
\begin{align}
    \overline{a}_\mathrm{ss} = \frac{i \sqrt{\kappa_\mathrm{c}} \beta_\mathrm{in}}{\Delta_\mathrm{c} - i \left[ \frac{\kappa}{2} + \frac{2 N_\mathrm{eff} (\overline{g}^2 + \sigma^2)}{\gamma + \Delta_\mathrm{inh}} \right]} ~,
    \label{eq:assbar}
\end{align}
i.e., the disorder-averaged self energy is purely imaginary and merely renormalizes the cavity decay rate. 
Here, $N_\mathrm{eff} < N$ is the effective number of ions in the ensemble that contribute to the sum in Eq.~\eqref{eq:ass}, i.e., those ions that have not been depolarized into a state with $s_z^j \approx 0$.
This results also reveals that the relevant collective cooperativity that controls renormalizations of the bare cavity damping rate $\kappa$ due to its interaction with the spin ensemble is $C_\mathrm{inh} = 4 N_\mathrm{eff} \overline{g}^2 / \Delta_\mathrm{inh} \kappa$, where we assumed $\sigma^2 \ll \overline{g}^2$ and $\gamma \ll \Delta_\mathrm{inh}$. 
Even though the collective cooperativity $C$ may be large, $C_\mathrm{inh} \approx 0.24$ is only modest due to the large inhomogeneous broadening $\Delta_\mathrm{inh}$. 
Surprisingly, this result provides an excellent description of the disorder-averaged steady-state cavity field even in the limit of strong driving, where spins resonant with the drive laser are almost depolarized, $s_z^j \to 0$. 
An intuitive explanation is the CIT effect \cite{Lei2023}: For $\vert \overline{a}_\mathrm{ss} \vert g_j \gg \gamma$, ions in a window $\vert \delta_j \vert \ll \vert \overline{a}_\mathrm{ss} \vert g_j$ around the drive laser frequency will be highly depolarized, $s_z^j \approx 0$, and their contributions to the self energy term in Eq.~\eqref{eq:ass} will be suppressed. 
Consequently, the self-energy will be dominated by spins with a large detuning  $\vert \delta_j \vert \gg \vert \overline{a}_\mathrm{ss}  \vert g_j$, but these spins have $s_z^j \approx -1$ such that the weak-drive assumption used to derive Eq.~\eqref{eq:assbar} remains valid if $N_\mathrm{eff}$ is suitably adjusted.
For our numerical parameters, the renormalization of the cavity decay rate by the self-energy term is less than $24\%$ due to the small value of $C_\mathrm{inh}$.

Given the analytical expression~\eqref{eq:assbar}, we decompose the cavity field into its semiclassical disorder-averaged steady-state value and quantum fluctuations $\hat{d}$, $\hat{a} = \overline{a}_\mathrm{ss} + \hat{d}$, and find that, to leading order, $\hat{H}$ decouples into a sum of $N$ independent single-spin Hamiltonians $\hat{H}_j = \delta_j \hat{\sigma}_z^j/2 + g_j (\overline{a}_\mathrm{ss}^* \hat{\sigma}_-^j + \overline{a}_\mathrm{ss} \hat{\sigma}_+^j)$, each describing a coherent Rabi drive.
This Rabi drive competes with single-spin relaxation and detuning $\delta_j$ such that each spin relaxes to a steady state with
\begin{align}
    s_{z,\mathrm{ss}}^j 
        &= -1 + \frac{8 \vert \overline{a}_\mathrm{ss} \vert^2 g_j^2}{8 \vert \overline{a}_\mathrm{ss} \vert^2 g_j^2 + \gamma^2 + 4 \delta_j^2}~, 
        \label{eq:szss1}\\
    s_{-,\mathrm{ss}}^j 
        &= - \frac{2 \overline{a}_\mathrm{ss} g_j}{8 \vert \overline{a}_\mathrm{ss} \vert^2 g_j^2 + \gamma^2 + 4 \delta_j^2} (i \gamma + 2 \delta_j)~.
        \label{eq:smss}
\end{align}
These expressions agree very well with the corresponding quantities obtained from a numerical solution of the semiclassical equations of motion~\eqref{eq:SC:a} to~\eqref{eq:SC:sz}.

\subsection{Fluorescence decay rate}
\label{sec:Theory:FluorescenceDecayRate}

We now analyze the fluorescent decay of the initial state given by Eqs.~\eqref{eq:ass}, \eqref{eq:szss1}, and~\eqref{eq:smss}.
Since the laser drive is switched off during the fluorescence measurement, the disorder-averaged fluoresence signal is given by
\begin{align}
    \overline{F(t)} 
    = \overline{\kappa_\mathrm{c} \vert a \vert^2} 
    = \frac{1}{n_\mathrm{traj}} \sum_{k=1}^{n_\mathrm{traj}} \kappa_\mathrm{c} \left\vert a^{(k)}(t) \right\vert^2 ~,
    \label{eq:Favg}
\end{align}
where $a^{(k)}(t)$ denotes the cavity field calculated numerically for the disorder realization $k \in \{1, \dots, n_\mathrm{traj}\}$. 
Examples of the simulated fluorescence curves are shown in Figure~\ref{fig:3}a and display two different regimes: 
Immediately after the end of the laser pulse, at time $t_\mathrm{pulse}$, the cavity is still populated with photons generated by the laser drive. 
These photons decay out of the cavity for times $t - t_\mathrm{pulse} \lesssim 10/\kappa$.
Because we wait $2\,\mu$s before opening the collection AOM, this regime is not observable in the experiments. 
At much longer times $t - t_\mathrm{pulse} \gg 10/\kappa$, the fluorescence is dominated by slow emission from the spin ensemble. 
This regime corresponds to the experimental measurements shown in Fig.~\ref{fig:2}c-d. 
To extract the relaxation rate of $\overline{F(t)}$ in this regime, we fit the numerically obtained fluorescence data to an exponential decay $c e^{- \Gamma (t - t_\mathrm{pulse})}$ with free parameters $c$ and $\Gamma$.

\begin{figure}
    \centering
    \includegraphics[width=0.4\textwidth]{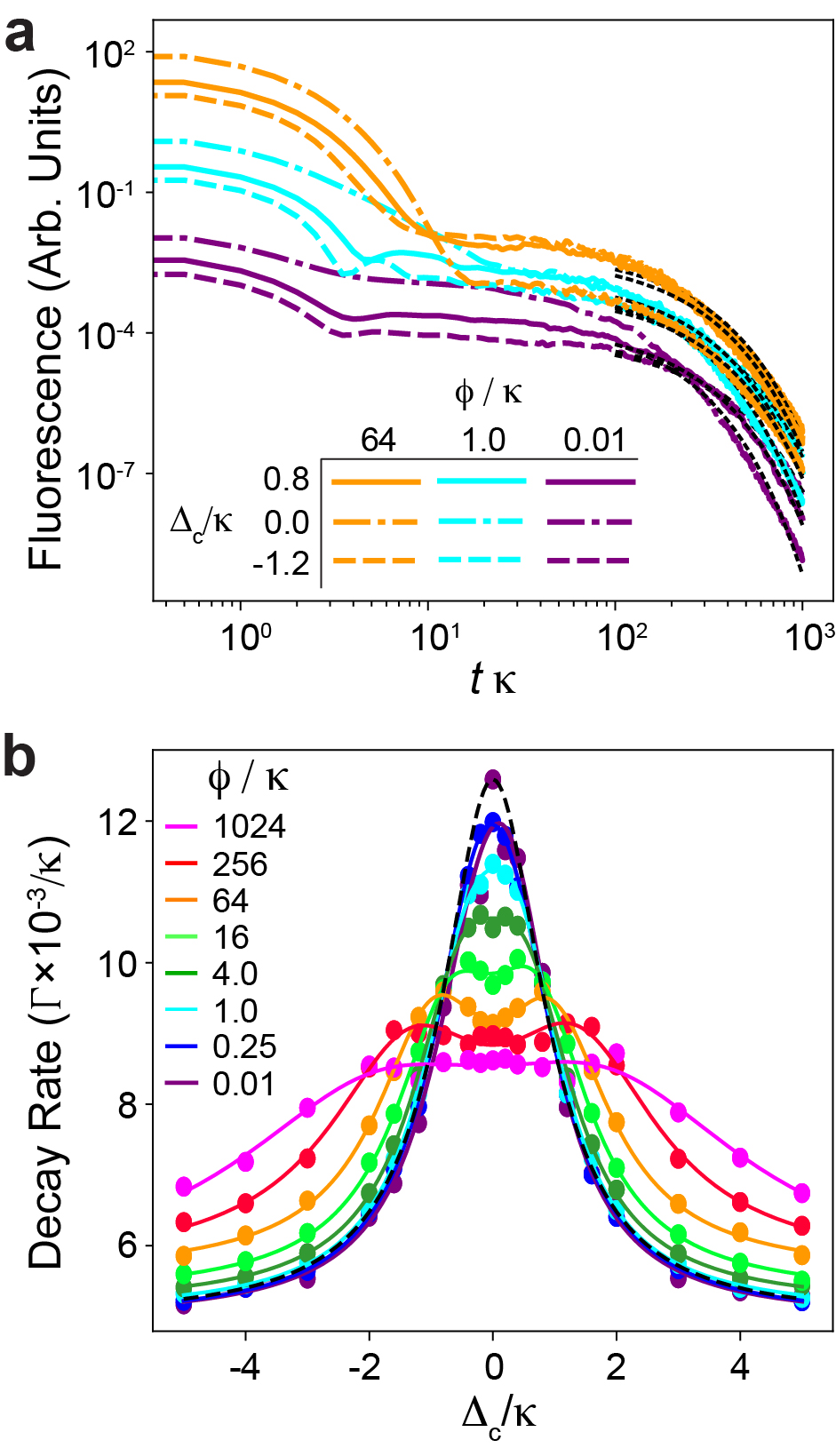}
    \caption{
    Simulated PLE measurements on cavity-coupled Er ensembles, based on the semiclassical model of the relaxation dynamics given by Eqs.~\eqref{eq:SC:a} to~\eqref{eq:SC:sz}.
    (a) Nine examples of the disorder-averaged fluorescence $\overline{F(t)}$, defined in Eq.~\eqref{eq:Favg}, for three different incident photon-flux values ($\phi = \vert \beta_\mathrm{in} \vert^2$) and cavity-laser detuning values $\Delta_\mathrm{c}$. 
    The various dotted black lines are exponential fits $c \exp(- \Gamma t)$ to the decay dynamics for $t \gg 1/\kappa$.
    (b) The spontaneous decay rate relative to the cavity decay rate ($\Gamma/\kappa$) as a function of the cavity-laser detuning ($\Delta_\mathrm{c}$) for increasing photon flux values ($\phi$). 
    The solid lines for each flux are double Lorentzian fits to the simulated data. 
    The dashed black line is a single Lorentzian function with FWHM $2 \kappa$.
    % \Martin{(c) Total fluorescence $\mathcal{F}$ after subtracting the decay of the initial photon field, defined in Eq.~\eqref{eq:TotalFluorescence}, as a function of the input flux $\phi$. The dashed fit function represents the PLE function~\eqref{eq:PLE} and is used to dermine the critical flux $\phi_0$ where saturation effects start to become relevant.}
    % The simulation parameters are $N=61$, $n_\mathrm{traj}=120$, $\Delta_\mathrm{inh} = 5 \kappa$, $\kappa_\mathrm{c} = 0.8 \kappa$, $\gamma=0.005\kappa$, $\overline{g}=0.07 \kappa$, $\sigma=0.01\kappa$, and $t_\mathrm{pulse} = 400/\kappa$.
    % Inset: Comparison between the semiclassical simulations and the experimental measurements. 
    % Maximum decay rate ratio (DRR) as a function of the incident flux $\phi$ relative to the corresponding value for the smallest incident flux, $\max_{\mathrm{\Delta}_\mathrm{c}} (\Gamma) / \max_{\mathrm{\Delta}_\mathrm{c}} (\Gamma \vert_{\phi_\mathrm{min}})$, as a function of the photon flux ($\phi$), normalized to the critical photon flux ($\phi_0$) where saturation effects start to occur (see main text). The gray shaded areas indicate the standard deviation of the experimental data.
    % \Martin{Add Fig.~\ref{fig:4}(b) here together with its caption text, remove curve ``local'' in Fig.~\ref{fig:4}(b).}
    }
    \label{fig:3}
\end{figure}

The ratio between the spontaneous decay rate $\Gamma$ and the cavity decay rate $\kappa$ is shown in Figure~\ref{fig:3}b.
Despite the fact that we approximate the full quantum dynamics with a semiclassical model and ignore the presence of Er ions coupled to the bus waveguide, the spontaneous decay rate qualitatively reproduces the experimental measurements shown in Figure~\ref{fig:1}d:
At the smallest input flux $\phi = 0.01 \kappa$, the spontaneous decay rate is maximum on resonance, $\Delta_\mathrm{c} = 0$ and decreases for finite cavity detuning according to a Lorentzian function whose FWHM is given by $2\kappa$, $L(\Delta_\mathrm{c},0,\kappa)$ (dashed black curve in Fig.~\ref{fig:3}b). 
This is in rough agreement with our experimental findings where the Lorentzian lineshape of the Purcell enhanced decay rate is $\approx 35$\% wider than the cavity linewidth as determined from reflection scans.
At a slightly larger input flux of $\phi = 0.25 \kappa$, the curve of the decay rate shows a flat region around zero detuning, which is consistent with the curve being the sum of two detuned Lorentzian functions. 
This interpretation also lets us understand the behavior at higher fluxes $\phi \gtrsim 16 \kappa$, where one sees two distinct local maxima at finite detuning $\vert \Delta_\mathrm{c} \vert > 0$.
The separation of these new maxima increases with drive power and ultimately exceeds $2 \kappa$.
At the same time, the decay rate for a resonant cavity decreases.

To quantify the separation between the two Lorentzian functions, we fit the decay rate to $a [ L(\Delta_\mathrm{c},\Delta_+,h_+) + L(\Delta_\mathrm{c},\Delta_-, h_-)] + b$. 
Free parameters are the overall amplitude $a$, the offset $b$, the detunings $\Delta_\pm$ of the two peaks, and the corresponding width parameters $h_\pm$. 
These fits are shown by the solid lines in Figure~\ref{fig:3}b.
In the SI, we compare the experimental results shown in Fig.~\ref{fig:2}d with their simulated equivalents shown in Fig.~\ref{fig:3}b using two figures of merit: 
The splitting $\Delta_+ - \Delta_-$ extracted from the Lorentzian fits and the ratio between the maximum decay rate $\Gamma$ at a given input flux $\phi$, $\max_{\Delta_\mathrm{c}} (\Gamma\vert_\phi)$, and its corresponding value at the smallest input flux $\phi_\mathrm{min}$, $\max_{\Delta_\mathrm{c}} (\Gamma\vert_{\phi_\mathrm{min}})$.
In both cases, we find that the semiclassical model agrees with the experimental data for $\phi \ll \phi_0$, where $\phi_0$ is the critical flux above which saturation effects start to become relevant (see SI).
Larger deviations between the semiclassical and the experimental data emerge when entering the saturation regime $\phi \gg \phi_0$.
%, the numerical simulations predict a slower increase of the frequency splitting with increasing $\phi$. 
We attribute this effect to the different values of the inhomogeneous broadening $\Delta_\mathrm{inh}$ in the experiment and the simulations:
Intuitively, $\Delta_\mathrm{inh}$ determines the spectral extent of the ion ensemble and defines an upper bound on the maximum possible splitting $\Delta_+ - \Delta_-$. 
Due to the smaller value of $\Delta_\mathrm{inh}$, these finite size effects of the ion ensemble will manifest themselves more pronouncedly in the simulations.

\subsection{Anomalous spin relaxation and toy model of the decay dynamics}

Both the experimental data and the semiclassical simulations
exhibit a striking transition from a weak-drive regime where $\Gamma$ decays monotonically with increasing $\vert \Delta_\mathrm{c} \vert$ (i.e., a single peak in the optical decay rate versus $\vert \Delta_\mathrm{c} \vert$, c.f.\ Figs.~\ref{fig:2}d and \ref{fig:3}b) to a high-drive regime where $\Gamma$ is non-monotonic with increasing $\vert \Delta_\mathrm{c} \vert$ (i.e., two symmetric peaks in $\Gamma$ at nonzero cavity detunings).  As we now discuss, the simulations provide insight into this phenomenon, as they show that a similar transition occurs if we look at the frequency-resolved decay of different spins in our ensemble.

To understand how spins with a certain detuning $\delta_j$ from the drive laser decay, we employ a binning procedure. 
For each disorder realization, we assign each spin to a frequency bin of width $0.167 \kappa$ based on their random detuning $\delta_j$.  
Averaging over many realizations, we can then compute the average time-dependent spin excitation $1 + s_z^j$ in each frequency-resolved bin.  
We present results for the case $\Delta_c = 0$ in the SI.  
Surprisingly, we find that for sufficient drive strengths, spins that are resonant with the cavity ($\delta_j \approx \Delta_\mathrm{c} = 0$) decay far slower than spins with a large detuning (see Fig.~S9 of the SI).
While this feature cannot be probed in the current experiment, it could be measured by analyzing the frequency-resolved fluorescence using a narrow spectral filter.
The slow decay of resonant spins is at odds with the naive expectation that the Purcell effect enhances the relaxation of resonant spins, which should therefore decay faster than detuned spins.
To provide an intuitive explanation of this effect, we now derive and analyze a simple toy model that captures the essential physics of this spin relaxation process and reproduces nonmonotonic decay behavior at large input fluxes.

The key observation is that the steady-state spin coherence at the end of the laser pulse [Eq.~\eqref{eq:smss}] is non-monotonic as a function of spin detuning $\delta_{j}$. Resonant spins have zero coherence, $s_{-,\mathrm{ss}}^j(\delta_j = 0) \approx 0$, due to the interplay between local decay and the Rabi drive generated by the cavity field. Similarly, the coherence of highly detuned spins tends to zero, $s_{-,\mathrm{ss}}^j(\vert \delta_j \vert \to \infty) \to 0$, as they remain in the ground state. At intermediate detunings, however, there is a group of spins whose coherence is non-negligible. We argue that this coherence allows these detuned spins to effectively drive the resonant spins via the cavity, an unusual kind of cavity-mediated spin-spin interaction.  This additional driving leads to both long-lived excitations in the resonant spins, and 
%to an amplified cavity output
%\Martin{is ``amplified'' the right word here? ``enhanced''?}
%at non-zero cavity detuning.
%\Aash{I'm also not entirely clear on what is meant here.  Just say something milder like "...and to a non-monotonic dependence of fluorescence decay on detuning"?}
%\Martin{fine with me}
to a non-monotonic dependence of fluorescence decay on detuning.
%the off-resonant spins to coherently pump their excitations back into the cavity during the decay dynamics, leading to both enhanced survival of the resonant spins, and to an amplified cavity output at non-zero cavity detuning.

\begin{figure}
    \centering
    \includegraphics[width=0.4\textwidth]{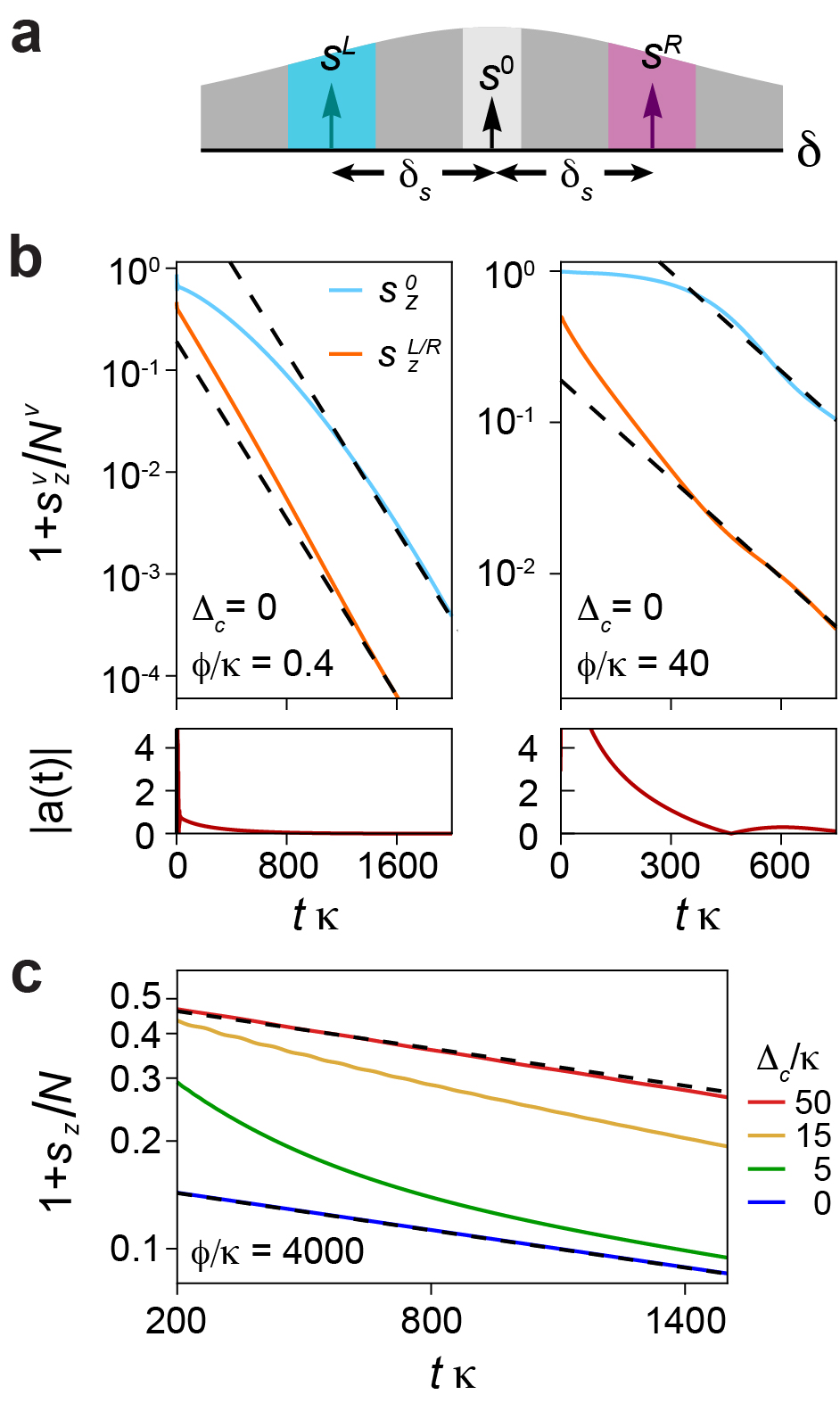}
    \caption{Relaxation dynamics for a toy three-spin model. (a) Schematic of the model, using one resonant spin $s^0$ and two off-resonant spins $s^{L/R}$ with detunings $\pm \delta_S$. (b) Decay dynamics of the spins and cavity for different initial pulse $\phi=0.4, 40$ for fixed $\gamma=0.005\kappa$, $g=0.004\kappa$, $N^{0}=2,000$, $N^{L/R} = 40,000$. Black dashed lines are intrinsic decay $\sim e^{-\gamma t}$.
    For a strong drive pulse, there is no appreciable short-time decay of $s^0_z$.  
    (c) Decay dynamics of all spins $s_z = s_z^L + s_z^0 + s_z^R$ normalized by total ion number $N = N^{0} + N^{L} + N^{R}$ at longer times $t\kappa\geq 200$. We use a smaller $\gamma=4\times 10^{-4}\kappa$ to accentuate transient features. One sees clearly that the effective decay rate of excitations has a non-monotonic dependence with $\Delta_c$.
    }
    \label{fig:4}
\end{figure}

To extract and s qimplify the core physics, we coarse grain the inhomogeneous spin ensemble, replacing it with just three large collective spins as depicted in Fig.~\ref{fig:4}a. One spin (size $N^0$) is resonant, representing spins that are depolarized, with spin operators $s_{\alpha}^{0}$ for $\alpha \in \{+,-,z\}$). The other two spins (sizes $N^L = N^R$) are taken to have equal and opposite fixed detunings $\pm \delta_S$, representing the set of off-resonant spins with maximal nonzero coherence, with spin operators $s_{\alpha}^{L}$, $s_{\alpha}^{R}$.  
By chosing $(N^L = N^R) \gg N^0 \gg 1$, we can use this ensemble of three large collective spins to qualitatively capture key aspects of the full ensemble. This hierarchy reflects the fact that there are many detuned spins with near optimal initial coherence (see SI for more details).      
%These three spins are meant to represent entire collections of spins with similar detuning; hence we take the spin sizes to be large, $N^0$, $N^{L}$, $N^{R} \gg 1$ with $N^{L} = N^{R}$ (see SI for details). We seek to condense all off-resonant spins in the ensemble into large spins with characteristic detunings that yield the strongest contributions to the experimental signal. The spin sizes are thus meant to qualitatively capture the number of ions; hence we consider $N^{L/R} \gg N^{0}$ due to inhomogeneous broadening, although the ensuing physics is weakly dependent upon the specific numbers chosen.

The relaxation dynamics for such a three-spin system are solved semi-classically, with initial conditions equivalent to the steady-state solutions of the laser-driven full ensemble. The cavity field is initialized to $a(0) = \overline{a}_{\mathrm{ss}}$ from Eq.~\eqref{eq:assbar}, neglecting the spin contributions in the denominator for simplicity. The spins are initialized to $s_{\alpha}^{0,L,R} = N^{0,L,R} s_{\alpha,\mathrm{ss}}^{0,L,R}$  from Eqs.~\eqref{eq:szss1} and~\eqref{eq:smss}, with $\delta_{0,L,R} = 0, -\delta_{S}, +\delta_{S}$ respectively. While the spin excitation fractions $s_{z}^{0}$, $s_{z}^{L/R}$ monotonically increase with larger laser power, the spin coherence $s_{-}^{L/R}$ of the off-resonant spins exhibits a maximum at an intermediate power where $|\overline{a}_{\mathrm{ss}}|\approx \delta_{S}/(\sqrt{2} g)$, provided that $g |\overline{a}_{\mathrm{ss}}| \gg \gamma$. The amplitude at this maximum is $|s_{-}^{L/R}|\approx N^{L/R}/(2\sqrt{2})$, with $s_{-}^{L}\approx s_{-}^{R}$, roughly independent of spin size. This gives a characteristic detuning $\delta_{S} = \sqrt{2}g |\overline{a}_{\mathrm{ss}}|$ at which spin coherence is maximal and non-negligible (again provided that $g |\overline{a}_{\mathrm{ss}}| \gg \gamma$). Note that $\delta_{S}$ is not fixed, but increases for stronger laser power (larger $\phi$). For different laser powers, groups of spins at different detunings in the ensemble have coherence and can pump the cavity. This picture relies upon the ensemble being inhomogeneously broadened, and breaks down once $\delta_{S}$ becomes larger than the width of the ensemble frequency distribution.

Figure.~\ref{fig:4}b plots the relaxation dynamics of the spins and the cavity in this toy model, using the same parameters as the ensemble semiclassical simulations (see Fig.~\ref{fig:3}) for a resonant cavity $\Delta_c = 0$. For strong initial laser power (larger $\phi$) we find a striking enhanced survival of the resonant spins, where they remain nearly depolarized over a long timescale before resuming intrinsic decay $\sim e^{-\gamma t}$. This slow initial relaxation matches the behaviour seen in the semiclassical disorder-averaged simulations, and can be understood as follows. First, note that after a strong drive pulse, the resonant spin on its own will simply undergo intrinsic loss at rate $\gamma$. At the level of semiclassics, it cannot have any cavity-mediated Purcell decay because it is depolarized (and hence has no coherence, $s_{-}^{0}\approx 0$), which prevents it from coherently trading excitations with the cavity.
Effective relaxation of the resonant spin is further impeded by a second effect involving the non-resonant spins $s^L, s^R$.  Since these spins do have appreciable initial coherence, they can pump their excitations into the cavity. In addition to a Purcell-enhanced decay of the off-resonant spins, this effective pumping also generates a transient cavity field that can directly drive the resonant spin $s^0$.  
This keeps the resonant spin excited provided the effective driving rate $\sim g |a|$ is stronger than $\gamma$. As seen in the figure, the expected decay of the resonant spins resumes as soon as the off-resonant spins stop undergoing Purcell-enhanced decay and the cavity field depletes.

At an intuitive level, the cavity-mediated effective drive of the resonant spins leads to a reduced spin decay rate at high pump powers for $\Delta_\mathrm{c} = 0$ (as found in the simulations of the full ensemble shown Fig.~\ref{fig:3}b), since any decay through the cavity (which is anyways small at the semiclassical level for a highly depolarized state) will be counteracted by the off-resonant spins driving the cavity. This simple picture can also be used to qualitatively understand the emergence of a non-monotonic dependence of relaxation physics on the cavity detuning $\Delta_c$. Figure~\ref{fig:4}c plots the relaxation of all spins $s_z = s_z^L + s_z^0+s_z^R$ at long timescales $t\kappa \geq 200$, using a strong initial pulse 
$\phi = 4000\kappa$ and a smaller decay rate $\gamma = 4\times 10^{-4}\kappa$ to better visualize transient features. 
%Excitation of all spins is shown for completeness, although the signal is dominated by off-resonant spins since their size is much larger. 
We find a characteristic Purcell enhancement that only occurs for intermediate values of $\Delta_c$. This happens because the decay of off-resonant spins into the cavity must be both non-negligible, and also persist over sufficiently long timescales. At small cavity detunings the off-resonant spins decay into the cavity very quickly, and are depleted before the chosen time window is reached. At larger cavity detunings,  Purcell-enhanced decay of the detuned spins is slowed, leading observable consequences over longer time scales. Going further to the limit $\Delta_{c} \to \infty$, any coherent coupling of spins to the cavity is energetically suppressed, and the spins are very weakly excited to begin with ($|\overline{a}_{\mathrm{ss}}|\to 0$ so $s_{-}^{L/R} \to 0$), hence no excitations can be transferred before the off-resonant spins decay due to their own intrinsic loss.

\section{Conclusion}

In this work, we have used an inhomogeneously broaden ensemble of Er ions evanescently coupled to a relatively narrow optical nanocavity to explore optical decay dynamics as a function of laser-cavity detuning and pump power. In order to realize these measurements, we developed a materials-compatible nanofabrication process to greatly improve the device-to-fiber photon collection efficiency. Our resonant PLE measurements reveal an interesting three-fold suppression of the Purcell-enhanced optical lifetime at the highest pump fluence measured when the cavity is resonant with the optical transition. We estimate that the number of ions participating in this effect is roughly 2000 at the highest incident photon flux. We have employed a semiclassical model that captures qualitative aspects of the experimental optical relaxation even with a modest number of $N=61$ ions.

From these simulations, there appear to be rich decay dynamics that happen in the time regime between traditional cavity photon and spin decay processes ($\sim 1/\kappa < t < 1/\Gamma$) that are not being accessed in the current experimental configuration. In the future, it would be beneficial to switch to gated operation of single photon detectors that can simultaneously enable higher collection efficiency by removal of a lossy modulator in the collection path and detection of photons earlier in the relaxation process. In addition, there are likely other interesting dynamics that occur as the photons impinge and reflect from the cavity, as has been explored recently in the regime of spectrally narrower but larger ensembles of ions coupled to lower Q cavities \cite{Lei2023}. In future experiments, it would be interesting to use an ultra narrowband optical filter the collected emission via PLE to to spectrally discern differences in emitter contribution to the PLE signal.

From a modeling perspective, it would be beneficial to optimize the numerical implementation to investigate these effects in larger ensembles that approach the experimental number of ions probed. A much larger simulated ensemble size can enable us to modify other parameters to better emulate experimental conditions: such as increasing the ensemble inhomogeneous linewidth, reducing the cavity-waveguide coupling efficiency $\kappa_\mathrm{c}$, and broadening the disorder averaged ion-cavity coupling strength. In doing so, we may be able to reduce potential finite-size effects in the splitting of the peak decay rates as a function of $\phi$ to better match experimental results. On the theory side, an analytical understanding of the emergence of the double-peaked structure in the fluorescence decay rate is desired, and such insights would provide clarity on the origin of the surprising survival of excitations in spins that are resonant with the cavity mode. 

\begin{backmatter}
\bmsection{Funding}
The authors acknowledge the Q-NEXT Quantum Center, a U.S. Department of Energy, Office of Science, National Quantum Information Science Research Center, under Award Number DE-FOA-0002253 for support (M. K., C. J., G. D. G., I. M., D. D. A., S. G., A. A. C., A. M. D.). Additional materials characterization support (M. T. S., S. E. S., and F. J. H.) was provided by the U.S. Department of Energy, Office of Science; Basic Energy Sciences, Materials Sciences, and Engineering Division.

\bmsection{Acknowledgments}
All electron microscopy and device fabrication were performed at the Center for Nanoscale Materials, a U.S. Department of Energy Office of Science User Facility, supported by the U.S. DOE, Office of Basic Energy Sciences, under Contract No. DE-AC02-06CH11357. The authors would like to thank D. Czaplewski, C. S. Miller, and R. Divan for assistance with device fabrication. 

\bmsection{Disclosures}

\noindent ``The authors declare no conflicts of interest.''

\bmsection{Data Availability} Data underlying the results presented in this paper are not publicly available at this time but may be obtained from the authors upon reasonable request.

\bmsection{Supplemental document}
See Supplementary Information for supporting content. 

\end{backmatter}

%%%%%%%%%% If using BibTeX:
\bibliography{sample}

\end{document}